\documentclass[12pt,a4paper]{article}
\usepackage{a4wide}
\usepackage{amsmath,amssymb,amsfonts,amsxtra,simpler-wick} 
\usepackage{cite}

\allowdisplaybreaks

\begin{document}
\thispagestyle{empty}


\vskip 3.0cm

\begin{center}{\Large \bf Quantum gravitational corrected evolution equations of charged black holes}\end{center}  

\vspace*{1cm}

\centerline{Ruben Campos Delgado\footnote{E-mail: ruben.camposdelgado@gmail.com}}

\vspace{1cm}

\begin{center}{\it
Bethe Center for Theoretical Physics,\\
Physikalisches Institut der Universit\"at Bonn,\\
Nussallee 12, 53115 Bonn, Germany}\\
\end{center}

\vspace*{1cm}

\centerline{\bf Abstract}
\vskip .3cm

We explain how quantum gravity, treated as an effective field theory, might modify the evaporative evolution of a four-dimensional, non-extremal, non-rotating, charged black hole.  With some approximations, we derive a set of coupled differential equations describing the charge and mass of the black hole as a function of time. These equations represent a generalisation of the analogous ones already present in the literature for classical black holes.


\vskip .3cm

\newpage
\section{Introduction}\label{sec:intro}
Black holes in general relativity never vanish and are eternal. The formalism of quantum field theory in curved spacetime implies however that black holes are complicated entities which emit thermal radiation and have an entropy proportional to the horizon area (Bekenstein-Hawking area law) \cite{Hawking:1976de,Bekenstein:1973ur}. Yet this is not the full story. It is generally expected that any complete theory of quantum gravity (like string theory or loop quantum gravity) should not only reproduce the Bekenstein-Hawking area law, but also corrections to it. Such corrections are both of logarithmic \cite{Das:2001ic, Pourhassan:2015cga} and exponential \cite{Chatterjee:2020iuf, Pourhassan:2020yei} nature.  The black hole stability is also affected: the black hole can stop evaporating and becomes stable \cite{Upadhyay:2019hyw}.

A convenient way to keep track of the quantum gravitational corrections consists in treating gravity as an effective field theory. One starts with the path integral for matter fields coupled to gravity, then integrates out the fluctuations of the matter fields, thereby obtaining an effective action which is an expansion in terms of powers of the curvature containing both local and non-local contributions. 
Effective field theory methods applied to black hole physics have recently drawn a renewed interest. Calmet and Kuipers \cite{Calmet:2021lny} computed the quantum gravitational corrections to the entropy of Schwarzschild black holes by applying the Wald formula \cite{Wald:1993nt} and integrating over the quantum corrected event horizon. A generalisation of this procedure for electrically charged (Reissner-Nordstr\"om) black holes was carried out shortly thereafter \cite{Delgado:2022pcc}. 

In this paper we continue in this direction and, within the effective field theory framework, we study quantum gravity effects on the evaporative evolution of a non-extremal Reissner-Nordstr\"om black hole. In particular, we derive the differential equations describing the loss of mass $M$ and charge $Q$. The difficulty of an exact calculation is insurmountable at the present stage. Thus, some approximations have to be taken into account. We assume that the black hole is sufficiently isolated so that accretion processes can be neglected. Accordingly, the evolution of the black hole is dominated by Hawking evaporation. Moreover, we work in the slow-evolution regime, where the mass and charge do not change significantly  on a geometrical time scale ($\tau\sim M$). This permits to describe the black hole by the (quantum corrected) Reissner-Nordstr\"om metric with $M$ and $Q$ being slowly varying functions of time. The differential equations that we derive represent a generalisation of those obtained, without taking quantum gravity into account, by Hiscock and Weems \cite{Hiscock:1990ex} at the end of the 1980's. 

The paper is organised as follows. In Section \ref{sec:qg} we review the procedure needed to obtain the quantum gravitational corrections to the  Reissner-Nordstr\"om metric. In Sections \ref{sec:charge} and \ref{sec:mass} we derive the laws of charge and mass loss, respectively. We summarise our results and discuss future directions in Section \ref{sec:conclusions}. Natural units $\hbar=c=k_B=1$ are used throughout the paper.
\section{Quantum gravitational corrections to the Reissner-Nordstr\"om geometry}\label{sec:qg}
Let us summarise few essential results concerning the quantum gravitational corrections to the geometry of black holes.  The starting point is an effective action containing both local and non local terms. At second order in curvature it reads \cite{Weinberg:1980gg, Starobinky:1981, Barvinsky:1983vpp, Barvinsky:1985an, Barvinsky:1987uw, Barvinsky:1990up, Donoghue:1994dn}
\begin{equation}\label{eq:local_action}
\begin{gathered}
    \Gamma=\int d^4x\, \sqrt{-g}\,\bigg(\frac{R}{16\pi G_N}+c_1(\mu)R^2
    +c_2(\mu)R_{\mu\nu}R^{\mu\nu}
    +c_3(\mu)R_{\mu\nu\rho\sigma}R^{\mu\nu\rho\sigma}\bigg)\\
    -\int d^4 x \sqrt{-g}\bigg[\alpha R\ln\left(\frac{\Box}{\mu^2}\right)R
    +\beta R_{\mu\nu}\ln\left(\frac{\Box}{\mu^2}\right)R^{\mu\nu} + \gamma R_{\mu\nu\rho\sigma}\ln\left(\frac{\Box}{\mu^2}\right)R^{\mu\nu\rho\sigma}\bigg],
\end{gathered}
\end{equation}
where $\mu$ is an arbitrary energy scale. The exact values of the coefficients $c_1$, $c_2$, $c_3$ can be obtained provided that one assumes an ultra-violet complete theory of quantum gravity, see for example \cite{Myrzakulov:2014hca, Elizalde:2017mrn}. On the other hand, the coefficients $\alpha,\beta,\gamma$ are calculable in a model independent way \cite{Donoghue:2014yha}. The logarithm appearing in the action is an operator with the integral representation \cite{Donoghue:2015nba}
\begin{equation}
    \ln\left(\frac{\Box}{\mu^2}\right)=\int_0^{+\infty}ds\, \left(\frac{1}{\mu^2+s}-\frac{1}{\Box+s}\right).
\end{equation}
When charged black holes are considered, the gravitational action is supplemented by the Maxwell action
\begin{equation}
    \Gamma_{M}=-\frac{1}{4}\int d^4x\,\sqrt{-g}\,F_{\mu\nu}F^{\mu\nu},
\end{equation}
where $F_{\mu\nu}=\partial_{\mu}A_{\nu}-\partial_{\nu}A_{\mu}$ is the electromagnetic tensor and $A_{\mu}$ is the electromagnetic potential.  In principle, one could add additional terms for the electromagnetic part. For instance,
\begin{equation}
     \Gamma_{M}=\int d^4x\,\sqrt{-g}\,\left(-\frac{1}{4} F_{\mu\nu}F^{\mu\nu}+d_1F^2R+d_2F^4+\cdots \right).
\end{equation}
To further simplify the calculations, we will neglect these terms for now on and set $d_i=0$.

Upon varying the full action $\Gamma+\Gamma_M$ and employing perturbation theory in the Newton constant $G_N$, one obtains a solution corresponding to a quantum gravitational corrected Reissner-Nordstr\"om black hole \cite{Delgado:2022pcc}:
\begin{equation}\label{eq:final_metric}
ds^2=-f(r)dt^2+\frac{1}{g(r)}dr^2+r^2d\theta^2+r^2\sin^2\theta d\phi^2,
\end{equation}
where
\begin{equation}
\begin{split}\label{eq:f(r)}
    f(r)=&1-\frac{2G_N M}{r}+\frac{G_N Q^2}{r^2}-\frac{32\pi G^2_N Q^2}{r^4} \Big[c_2+4c_3\\&+\left(\beta+4\gamma\right)\left(2\ln\left(\mu r\right)+2\gamma_E-3\right)\Big]+\mathcal{O}(G^3_N),
\end{split}
\end{equation}
\begin{equation}
\begin{split}\label{eq:g(r)}
   g(r)=&1-\frac{2G_N M}{r}+\frac{G_N Q^2}{r^2}-\frac{64\pi G^2_N Q^2}{r^4}\Big[c_2+4c_3\\&+2\left(\beta+4\gamma\right)\left(\ln\left(\mu r\right)+\gamma_E-2\right)\Big]+\mathcal{O}(G^3_N),
\end{split}
\end{equation}
and $\gamma_E$ is the Euler-Mascheroni constant.
Moreover, not only the classical metric but also the non-vanishing independent component of $F_{\mu\nu}$ (and hence the electric field $E$) receives quantum gravitational corrections \cite{Delgado:2022pcc}:
\begin{equation}\label{eq:Ftr}
\begin{split}
    E(r)=&F_{tr}=-F_{rt}=\frac{Q}{r^2}+\frac{16\pi G^2_N Q^3}{r^6}\Big[c_2+4c_3\\&+\left(\beta+4\gamma\right)\left(2\ln\left(\mu r\right)+2\gamma_E-5\right)\Big]+\mathcal{O}(G^3_N).
\end{split}
\end{equation}
Although the above results seem to depend on the arbitrary energy scale $\mu$, the renormalised constants also carry an explicit scale dependence \cite{El-Menoufi:2015cqw}:
\begin{equation}\label{eq:RG_coefficients}
\begin{split}
    c_1(\mu)=c_1(\mu_*)-\alpha\ln\left(\frac{\mu^2}{\mu^2_*}\right),\\
    c_2(\mu)=c_2(\mu_*)-\beta\ln\left(\frac{\mu^2}{\mu^2_*}\right),\\
    c_3(\mu)=c_3(\mu_*)-\gamma\ln\left(\frac{\mu^2}{\mu^2_*}\right),\\
\end{split}
\end{equation}
where $\mu_*$ is some fixed scale where the effective theory is matched onto the full theory. Inserting Eq. \eqref{eq:RG_coefficients} into Eqs. \eqref{eq:f(r)}, \eqref{eq:g(r)} and \eqref{eq:Ftr}, one sees that the terms involving $\mu$ cancel out. 

The radius of the event horizon is a solution of the equation $g(r)=0$. For a non-critical black hole with $Q^2<GM^2$, the radius can be written as a power series in the electric charge. Up to order $Q^2$ it is \cite{Delgado:2022pcc}
\begin{equation}\label{eq:radius}
\begin{split}
    r_{+}=2G_NM - \frac{Q^2}{2M}+ 
     \frac{8\pi Q^2}{G_NM^3}\left[c_2+4c_3+2(\beta+4\gamma)\left(\ln(2G_NM\mu)+\gamma_E-2\right)\right].
\end{split}
\end{equation}
If the black hole is critical, $Q^2=GM^2$, then the quantum corrections in $g(r)$ appear only at order $\mathcal{O}\left(G^3_N\right)$, so the radius can be taken to coincide with the classical result, $r_{+}=G_NM$. Hence, the horizon radius is approximately bounded by $G_NM<r_+<2G_NM$. 
\section{Charge loss rate}\label{sec:charge}
In this section we derive a differential equation that approximately describes the loss of electric charge of the black hole. We assume that its radius is much larger that the reduced Compton wavelength of the electron, $r_{+}>>1/m=2 \text{MeV}^{-1}$. If this is the case, then electron-positron pairs are created by the electric field of the black hole by Schwinger effect \cite{Schwinger:1951nm} and the process is well described by ordinary flat-space quantum electrodynamics. The particles with charge of the same sign as that of the black hole are repelled towards infinity, while the particles with opposite charges are attracted and contribute to neutralise the overall charge of the black hole. If the electric field is much smaller than the critical value $E_c=m^2/e$, then the rate of pair creation per unit four volume is \cite{Gavrilov:1996pz, Xu:2019wak}
\begin{equation}\label{eq:rate}
    \frac{dN}{dtdV}=\frac{m^4}{4\pi^3}\left(\frac{E}{E_c}\right)^2e^{-\frac{\pi E_c}{E}}.
\end{equation}
For the next calculations it is convenient to define for brevity
\begin{equation}
\begin{split}
    \Omega(r)\equiv a+b\ln(\mu r)= 32\pi(\beta+4\gamma)\ln(\mu r)\\+16\pi\left[c_2+4c_3+(\beta+4\gamma)(2\gamma_E-5)\right].
\end{split}
\end{equation}
The condition $E(r_+)<<E_c$ translates into a lower bound for the black hole mass. $E(r_+)$ is maximal for the extremal case, thus
\begin{equation}\label{eq:inequality}
    E(r_+)=\frac{Q}{r^2_+}+\frac{G^2_N Q^3}{r^6_{+}}\Omega(r_+)=\frac{M^2G_N+\Omega(r_+)}{M^3G^{5/2}_N}<<\frac{m^2}{e}.
\end{equation}
The coefficients $c_i$ are believed to be of order $O(1)$ \cite{Donoghue:1994dn}, while $\alpha$ and $\beta$ can reach values up to order $\mathcal O(100)$ \cite{Donoghue:2014yha}. If the mass of the black hole is much larger than the Planck mass, the factor $\Omega$ in Eq. \eqref{eq:inequality} can be neglected and the inequality becomes
\begin{equation}\label{eq:bound_on_mass}
    M>>\frac{e}{m^2 G^{3/2}}\sim 10^6 M_{\odot}.
\end{equation}
Hence, the formula for the pair creation rate keeps its validity as soon as the mass of the black hole is sufficiently large to obey the condition \eqref{eq:bound_on_mass}.

The charge loss rate of the black hole is obtained by integrating \eqref{eq:rate} over the entire space outside the horizon. We have
\begin{equation}
\begin{gathered}\label{eq:dQdt_first}
    \frac{dQ}{dt}=-e\frac{dN}{dt}=-\frac{e^3}{4\pi^3}\int_0^\pi d\theta\, \cos\theta \int_0^{2\pi} d\phi \\\times \int_{r_+}^{+\infty} dr\,r^2\left(\frac{Q}{r^2}+\frac{G^2_N Q^3\Omega}{r^6}\right)^2\exp\left\{\frac{-\pi m^2}{e\left(\frac{Q}{r^2}+\frac{G^2_N Q^3\Omega}{r^6}\right)}\right\}.
\end{gathered}
\end{equation}
Since the mass of the black hole is large, the horizon radius is large too. Therefore, the radial integral in Eq. \eqref{eq:dQdt_first} can be written as series expansion in $1/r_+$. The first non-trivial contribution is
\begin{equation}
\begin{gathered}
    \frac{dQ}{dt}=-\frac{e^3Q^2}{\pi^2}\int_{r_+}^{+\infty}dr\, e^{-\frac{\pi m^2 r^2}{eQ}}\bigg\{\frac{1}{r^2}+\frac{\pi m^2 Q G^2_N}{e}\frac{1}{r^4}\left[a+b\ln(\mu r)\right]+\mathcal{O}\left(r^{-6}\right)\bigg\}.
\end{gathered}
\end{equation}
The first integral is
\begin{equation}\label{eq:1integral}
\begin{gathered}
    \int_{r_+}^{+\infty}dr\, \frac{e^{-\frac{\pi m^2 r^2}{eQ}}}{r^2}=-\frac{\pi m}{\sqrt{eQ}}+\frac{e^{-\frac{\pi m^2 r^2_+}{eQ}}}{r_+}
    +\frac{\pi m}{\sqrt{eQ}}\text{erf}\left(\frac{\sqrt{\pi}mr_+}{\sqrt{eQ}}\right).
\end{gathered}
\end{equation}
Because of the condition \eqref{eq:bound_on_mass}, the argument of the error function is large and the function is well approximated by its asymptotic series. In this case, Eq. \eqref{eq:1integral} becomes
\begin{equation}
    \int_{r_+}^{+\infty}dr\, \frac{e^{-\frac{\pi m^2 r^2}{eQ}}}{r^2}\approx \frac{eQ}{2\pi m^2r^3_+}e^{-\frac{\pi m^2 r^2_+}{eQ}}.
\end{equation}
The second integral is
\begin{equation}
\begin{gathered}
    \int_{r_+}^{+\infty}dr\, \frac{e^{-\frac{\pi m^2 r^2}{eQ}}}{r^4}\left[a+b\ln(\mu r)\right]\\
    =\frac{1}{12} \Bigg\{3 \pi ^{3/2} b \left(\frac{m^2}{e Q}\right)^{3/2} G_{2,3}^{3,0}\left(\frac{\pi m^2
   r^2_+}{e Q}\bigg\lvert
\begin{array}{c}
 1,1 \\
 -\frac{3}{2},0,0 \\
\end{array}
\right)\\
-8 \pi ^2 \left(\frac{m^2}{e Q}\right)^{3/2} \left[a+b \log (\mu )\right] \text{erf}\left(\frac{\sqrt{\pi}m r_+}{\sqrt{eQ}}\right)\\+\frac{4}{r^3_+} \left[a+b \log (\mu )\right]
   e^{-\frac{\pi  m^2 r^2_+}{e Q}}+8\pi^2\left(\frac{m^2}{e Q}\right)^{3/2} \left[a+b \log (\mu )\right]\\-\frac{8\pi m^2}{eQr_+}\left[a+b \log (\mu )\right]e^{-\frac{\pi  m^2 r^2_+}{e Q}}\\+6 \pi ^{3/2} b
   \log (r_+) \left(\frac{m^2}{e Q}\right)^{3/2} \Gamma \left(-\frac{3}{2},\frac{\pi m^2   r^2_+}{e Q}\right)\Bigg\},
\end{gathered}
\end{equation}
where $\Gamma$ is the incomplete gamma function and $G_{2,3}^{3,0}$ is the Meijer G-function, defined as 
\begin{equation}
\begin{split}
    &G_{p,q}^{\,m,n} \!\left(\, z \, \Bigg\lvert \, \begin{matrix} a_1, \dots, a_p \\ b_1, \dots, b_q \end{matrix} \; \,  \right)  = \frac{1}{2 \pi i} \int ds\, \frac{\prod_{j=1}^m \Gamma(b_j - s) \prod_{j=1}^n \Gamma(1 - a_j +s)} {\prod_{j=m+1}^q \Gamma(1 - b_j + s) \prod_{j=n+1}^p \Gamma(a_j - s)} \,z^s.
\end{split}
\end{equation}
Approximating again the functions by their asymptotic series, we get
\begin{equation}
\begin{gathered}
    \int_{r_+}^{+\infty}dr\, \frac{e^{-\frac{\pi m^2 r^2}{eQ}}}{r^4}\left[a+b\ln(\mu r)\right]\approx \frac{eQ G^2_N}{2\pi m^2 r^5_+}\left[a+b\ln\left(\mu r_+\right)\right]e^{-\frac{\pi m^2 r^2_+}{eQ}}.
\end{gathered}
\end{equation}
Finally, we obtain the leading order terms of the differential equation governing the charge loss:
\begin{equation}\label{eq:charge_loss}
\begin{gathered}
    \frac{dQ}{dt}\approx -\frac{e^4}{2\pi^3m^2}\frac{Q^3}{r^3_+}e^{-\frac{\pi m^2r^2_+}{eQ}}\\
    -\frac{8e^3G_N^2}{\pi}\frac{Q^4}{r^5_{+}}e^{-\frac{\pi m^2r^2_+}{eQ}}\Big[c_2+4c_3
    +(\beta+4\gamma)\left(2\ln(\mu r_+)+2\gamma_E-5\right)\Big].
\end{gathered}
\end{equation}
For a non-extremal black hole the ration of the charge to the mass is small, so one can use Eq. \eqref{eq:radius} and write Eq. \eqref{eq:charge_loss}  as a series expansion in the charge (or the mass). At leading order,
\begin{equation}\label{eq:charge_loss_rate_final}
\begin{gathered}
    \frac{dQ}{dt}=-\frac{e^4Q^3}{16\pi^3m^2G^3_NM^3}e^{-\frac{4\pi m^2G^2_NM^2}{eQ}}\\
    -\frac{e^3 Q^4}{4\pi G^3_NM^5}e^{-\frac{4\pi m^2G^2_NM^2}{eQ}}\Big[c_2+4c_3\\
    +(\beta+4\gamma)\left(2\ln(2G_NM\mu )+2\gamma_E-5\right)\Big]+\mathcal{O}\left(\frac{Q^5}{M^5}\right).
\end{gathered}
\end{equation}
\section{Mass loss rate}\label{sec:mass}
The emission of Hawking radiation is the first mechanism with which a black loses mass. 
If the mass is large, then the temperature is low. Accordingly, all the mass lost in the Hawking process is due to the emission of massless particles. The mass loss rate due to the emission of black body radiation is given by the Stefan-Boltzmann law \cite{Hiscock:1990ex}:
\begin{equation}\label{eq:boltzmann}
    \frac{dM_1}{dt}=-\xi T^4\left(\frac{21}{8}\sigma_{\nu}+\sigma_{\gamma}+\sigma_g\right),
\end{equation}
where $\xi=\pi^2/60$ is the Stefan-Boltzmann constant and $\sigma_{\nu}$, $\sigma_{\gamma}$, $\sigma_g$ are the thermally averaged cross sections of the black hole for neutrinos, photons and gravitons, respectively. We neglect hereafter the mass of the neutrinos. 
It is convenient to introduce the geometrical optics cross section for the Reissner-Nordstr\"om black hole, denoted as $\sigma_0$. Defining
\begin{equation}
    \rho=\left(\frac{21}{8}\sigma_{\nu}+\sigma_{\gamma}+\sigma_g\right)\sigma^{-1}_0,
\end{equation}
Eq. \eqref{eq:boltzmann} can be rewritten as
\begin{equation}
    \frac{dM_1}{dt}=-\xi T^4 \rho \sigma_0.
\end{equation}
The classical value of $\sigma_0$ is \cite{Hiscock:1990ex, Crispino:2009ki}
\begin{equation}
    \sigma_0=\pi \frac{\left[3M+(9M^2-8Q^2)^{1/2}\right]^4}{8\left[3M^2-2Q^2+M(9M^2-8Q^2)^{1/2}\right]}.
\end{equation}
Let us now derive the quantum gravitational corrections to the classical result. 
The emitted particles move along null geodesics with equation
\begin{equation}
    \left(\frac{dr}{d\lambda}\right)^2=\frac{g(r)}{f(r)}\mathcal{E}^2-g(r)\frac{J^2}{r^2},
\end{equation}
where $\lambda$ is the affine parameter and $\mathcal{E}$, $J$ are the energy and angular momentum, respectively. For the emitted particles to reach infinity rather than falling back into the black hole horizon, one must require $(dr/d\lambda)^2\geq 0$, i.e.
\begin{equation}
    \frac{1}{l^2}\equiv\frac{E^2}{J^2}\geq \frac{f(r)}{r^2}
\end{equation}
for any $r\geq r_{+}$. The inequality is saturated by the maximal value of $f(r)/r^2$. The solution of
$\partial_r f(r)/r^2=0$ corresponds to the critical, unstable  orbit $r_c$ and the impact factor can be computed as $l_c=r_c/\sqrt{f(r_c)}$ \cite{Xu:2019wak, Xu:2019krv}.
The geometrical optics cross section of the black hole is then
\begin{equation}\label{eq:cross_section}
    \sigma_0=\pi l^2_c=\pi\frac{r^2_c}{f(r_c)}. 
\end{equation}
For a non critical black hole, we write $r_c$ as a power series in $Q$ (similarly to what done in Eq. \eqref{eq:radius}). Using the expression \eqref{eq:f(r)} for $f(r)$ we find
\begin{equation}
\begin{split}
    &\sigma_0=27\pi G^2_NM^2-9\pi G_NQ^2+\frac{32\pi^2 Q^2}{M^2}\Big[c_2+4c_3\\&
    +(\beta+4\gamma)\left(2\ln(3G_NM\mu)+2\gamma_E-3\right)\Big]+\mathcal{O}\left(Q^4\right).
\end{split} 
\end{equation}
The cross sections for neutrinos, photons and gravitons can be estimated as \cite{Hiscock:1990ex}
\begin{equation}
    \sigma_\nu\approx 0.66852 \sigma_0, \hspace{2 mm} \sigma_{\gamma}\approx 0.24044\sigma_0, \hspace{2mm} \sigma_g\approx 0.02748 \sigma_0.
\end{equation}
meaning that most of the power is emitted in neutrinos. The value of $\rho$ is then $\rho\approx 2.0228$.
The quantum gravitational corrected temperature of a non critical Reissner-Nordstr\"om black hole, up to order $\mathcal{O}(Q^2)$, is \cite{Delgado:2022pcc}
\begin{equation}
\begin{gathered}
    T=\frac{1}{8\pi G_N M}+\frac{Q^2}{4G^3_N M^5}\Big[2(c_2+4c_3)\\+\beta(4\gamma_E-9)+4\gamma(4\gamma_E-9)+4(\beta+4\gamma)\ln(2G_NM\mu)\Big].
\end{gathered}
\end{equation}
The contribution to the mass loss rate due to Hawking radiation is then
\begin{equation}
\begin{gathered}
    \frac{dM_1}{dt}=-\xi\rho\bigg\{\frac{27}{4096\pi^3G^2_NM^2}-\frac{9Q^2}{4096\pi^3G^3_NM^4}
    \\+\frac{Q^2}{512\pi^2G^4_NM^6}\Big[58(c_2+4c_3)+(\beta+4\gamma)\Big(116\gamma_E-255\\
    +108\ln(2G_NM\mu)+8\ln(3G_NM\mu)\Big)\Big]\bigg\}+\mathcal{O}\left(\frac{Q^4}{M^8}\right).
\end{gathered}
\end{equation}

The second mechanism of mass loss is due to the electric potential energy carried away by the repelled particles created by Schwinger effect. Assuming that the black-hole has a charge-to-mass ratio greater than $m/e$, then the potential energy gained by the repelled particle is greater than the rest mass of the electron. Scattering processes are then negligible and the motion of the created particles is dominated by the repulsive force. As a consequence, we approximate the mass loss rate due to the electromagnetic pair creation as the product of the potential energy lost per pair and the charge loss rate:
\begin{equation}
    \frac{dM_2}{dt}=\frac{dQ}{dt}\int_{r_+}^{+\infty}dr\, E(r).
\end{equation}
For a non-critical black hole,
\begin{equation}
    \frac{dM_2}{dt}=-\frac{e^4Q^4}{32\pi^3m^2G^4_NM^4}e^{-\frac{4\pi m^2 G^2_NM^2}{eQ}} +\mathcal{O}\left(\frac{Q^5}{M^6}\right).
\end{equation}
Finally, the leading contribution to the total mass loss rate is
\begin{equation}\label{eq:mass_loss_rate_final}
\begin{gathered}
    \frac{dM}{dt}=\frac{dM_1}{dt}+\frac{dM_2}{dt}=\\
    -\xi\rho\bigg\{\frac{27}{4096\pi^3G^2_NM^2}-\frac{9Q^2}{4096\pi^3G^3_NM^4}
    \\+\frac{Q^2}{512\pi^2G^4_NM^6}\Big[58(c_2+4c_3)+(\beta+4\gamma)\Big(116\gamma_E-255\\
    +108\ln(2G_NM\mu)+8\ln(3G_NM\mu)\Big)\Big]\bigg\}\\-\frac{e^4Q^4}{32\pi^3m^2G^4_NM^4}e^{-\frac{4\pi m^2 G^2_NM^2}{eQ}} +\mathcal{O}\left(\frac{Q^4}{M^8}, \frac{Q^5}{M^6}\right).
\end{gathered}
\end{equation}
\section{Conclusions and outlook}\label{sec:conclusions}
The main new results are given by the coupled differential equations \eqref{eq:charge_loss_rate_final} and \eqref{eq:mass_loss_rate_final} which respectively describe the approximate effect of quantum gravity on the evolution of charge and mass of a non-extremal, non-rotating, charged black hole for which accretion phenomena are negligible. In the limit where the quantum gravity effective action reduces to the usual Einstein-Hilbert action, one recovers the classical results of Hiscock and Weems \cite{Hiscock:1990ex}. Only the first non-trivial contributions arising from quantum gravity effects have been considered, but it is of course possible to expand the equations up to a generic power of charge and mass. These additional terms would be however even more suppressed.

The new equations can be taken as the starting point for future projects where more complicated scenarios are considered. For example, the equations are applicable in the late evolution of rotating (Kerr-Newman) black black holes, since these tend to lose relatively quickly  angular momentum \cite{Page:1976ki}.
One could also try to replicate the calculations presented in the paper for Anti-de Sitter black holes. In fact, the AdS$_2$ spacetime represents the near-horizon region of four (and higher-dimensional), charged black holes. Such black holes have a typical energy scale given by the AdS$_2$ radius, which is proportional to the charge, $l_{\text{AdS}}\sim Q$ \cite{Sarosi:2017ykf}. Now,  our equation \eqref{eq:charge_loss_rate_final} tells us that $Q$ is a function of time. This implies that the AdS$_2$ radius varies over time too. It would be interesting to investigate which consequences this fact may have on the AdS/CFT correspondence. Since it is impossible to solve our equations in closed-form, a numerical approach is needed in order to obtain a qualitative behaviour of $Q(t)$ and $M(t)$, which in turn could tells us more about the black hole instability and the possibility of recreating a holographic superconductor. 

Finally, it is worth stressing the fact that the equations were derived by applying several approximations. One could and should try to relax some initial assumptions, for example by going beyond the tree level Maxwell action. We expect to address some of these open questions in future publications. 

\section*{Acknowledgements}
This work was supported by the Bonn-Cologne Graduate School for Physics and Astronomy (BCGS).


\bibliographystyle{utphys}
\bibliography{references}

\end{document}